\documentclass[reprint,NumberedRefs]{JASA-EL}

\usepackage{amsmath,amssymb,amsfonts}
\usepackage{graphicx,color}
\usepackage{textcomp}
\usepackage{xcolor}
\usepackage[T1]{fontenc}
\usepackage[utf8]{inputenc}
\usepackage{microtype}
\usepackage{inconsolata}
\usepackage{booktabs}
\usepackage{multirow}
\usepackage{array}
\usepackage{makecell}
\usepackage{paralist}
\usepackage{url}
\makeatletter
\g@addto@macro\UrlBreaks{\UrlOrds}
\makeatother

\newif\ifrevnotes \revnotesfalse

\begin{document}

\title[JASA-EL/OTPM: A training-free diagnostic for SLD probes]{Off-target posterior mass: A training-free diagnostic for fine-tuning spoken language diarization probes}


\author{Zhuoyan Tao}
\email{zhuoyantao@gmail.com}
\correspondingauthor
\thanks{Current address: Language Technologies Institute, Carnegie Mellon University, Pittsburgh, Pennsylvania 15213, USA.}
\thanks{Submitted to JASA Express Letters.}
\author{Aditya Kommineni}
\email{akommine@usc.edu}
\author{Shrikanth Narayanan}
\email{shri@usc.edu}
\affiliation{Signal and Image Processing Institute, Ming Hsieh Department of Electrical and Computer Engineering, University of Southern California, Los Angeles, California 90089, USA}

\date{\today}

\begin{abstract}
Spoken Language Diarization (SLD) models provide frame-level language information of a speech recording. SLD modeling pipeline includes feature extraction from pre-trained speech models followed by clustering. Fine-tuning pre-trained speech models has shown performance gains for SLD with its magnitude being dataset dependent. \textbf{Off-target Posterior Mass (OTPM)}, computed based on posterior language identification probabilities is proposed as a diagnostic metric to forecast performance gains from SLD fine-tuning. OTPM obtains positive correlation with probe gains, 0.56 on unseen synthetic bilingual data and 0.23 on real-world corpora. These results indicate promising potential of OTPM as a diagnostic metric for SLD.
\end{abstract}


\maketitle

\section{Introduction}
\label{sec:intro}

\emph{Spoken Language Diarization} (SLD) models produce time-stamped segments for distinct languages in a given speech recording.\citep{mishra2024implicit, displace2024}
SLD models are vital for multilingual speech processing and facilitate downstream processing of utterances in code-switched and code-mixed contexts, that occur often in real-world settings.
While SLD is analogous to speaker diarization, it has idiosyncratic challenges: (1) languages within a family share acoustic and phonotactic structure, and code-switching produces
overlapping segments with no well-defined acoustic boundary\citep{wang2024bridging} (2) real-world corpora are often imbalanced, with one or two languages contributing a large portion of speech and the rest appearing as short, sparse turns.\citep{liu2025aligning, frost2022fine}

Most SLD systems comprise two stage pipelines.\citep{displace2024, sageld2025, taltech2024} 
Embedding/posterior Language Identification (LID) vector extraction from a pre-trained speech model (such as Whisper\citep{radford2023whisper} or MMS\citep{pratap2023mms}) for overlapping speech segments of a speech recording. 
Following this, the embeddings/posterior LID vectors are clustered (often using Agglomerative Hierarchical Clustering\citep{mishra2024implicit, sell2018diarization, park2020auto}) into the underlying languages in the recording.
Other methods for SLD include change-point detection methods\citep{mishra2021spoken} with attention and temporal smoothing, supervised frame-level classification\citep{wang2024bridging} and End-to-End systems such as SAGE-LD.\citep{sageld2025, mishra2023e2e, frost2022fine}
Evaluation of SLD models for the generated clusters are performed with \emph{language diarization error rate} (LDER) metric, computed similar to speaker diarization error rate with language labels in place of speaker labels (Eq.~\eqref{eq:lder}). 
Fine-tuning the pre-trained model embeddings/posterior LID vectors has been found to be an effective method for improving SLD performance in prior works.\citep{taltech2024}
Here, we will refer to this fine-tuning as training a \emph{probe} and the difference in LDER between the pre-trained embedding and fine-tuned embedding clustering is referred to as \emph{probe gain}.

While majority of prior work in the domain of SLD focuses on improving absolute diarization performance, in this work we probe whether it would be feasible to forecast the probe gain obtained from fine-tuning with a collected corpora, and we would want this forecasting to be in a training-free manner.
Being able to predict whether a given corpus would provide noticeable improvements on SLD task for a pre-trained speech model is valuable indicator for data curation without requiring explicit training, and would mark as an indicator of whether annotations should be performed.
Prior works in the domain of vision and language have explored similar questions, in the context of identifying pre-trained embedding quality through metrics such as NCE,\citep{tran2019nce} H-score,\citep{bao2019hscore} LEEP\citep{nguyen2020leep} and LogME,\citep{you2021logme} which score a pre-trained model's outputs or features against the \emph{labelled} target data, typically to rank candidate pre-trained models.
Alternatively, label-free accuracy prediction methods do not require labelled data, such as in the case of AutoEval by regressing accuracy on distribution shift across synthetic test sets,\citep{deng2021autoeval} and Average Threshold Confidence\citep{garg2022atc} by learning a threshold for model confidence.
Our proposed Off-Target Posterior Mass (OTPM) metric is in a training-free and label-free accuracy prediction paradigm, and uses the posterior LID probabilities from pre-trained speech models to compute proportion of probability assigned to non-target languages for the SLD task.
We show that this metric is correlated to probe gain obtained for SLD task on different synthetically generated and real-world datasets. 
The contributions of this work are as follows:
\begin{itemize}
    \item We propose Off-Target Posterior Mass (OTPM) as a metric for forecasting the probe gains obtained for Spoken Language Diarization task on speech foundation models.
    \item On an unseen set of 19 synthetically generated bilingual language pair data, OTPM obtains a Spearman correlation of 0.56 and 0.69 with probe gains for Whisper base and large-v3 models respectively. 
    \item When translating to real-world corpora, while OTPM is able to better align with probe gain compared to baselines, the correlation is smaller with 0.23 and 0.27 for Whisper base and large-v3 models.
\end{itemize}

\begin{figure*}[t]
  \centering
  \includegraphics{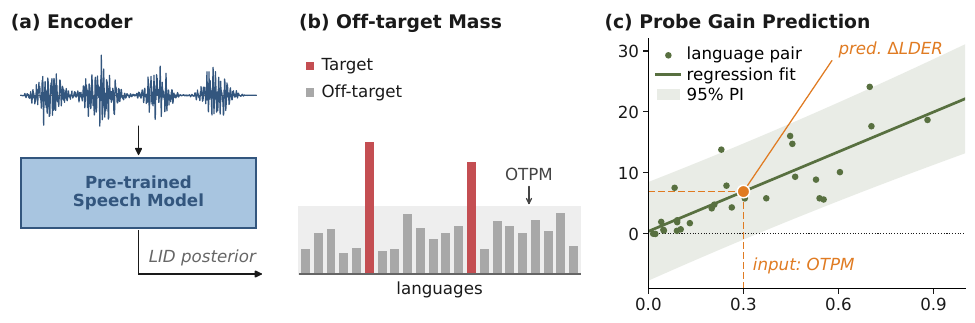}
  \caption{Off-target LID posterior mass (OTPM) as a training-free diagnostic for SLD
probe gain. \textbf{(a)} Each short window of the corpus audio is passed
through a pre-trained speech model, which produces a multilingual LID posterior.
\textbf{(b)} Posterior mass for Target languages ($\mathcal{T}$ marked in red) carry most of the mass in majority of cases;
OTPM is the fraction assigned to non-target languages ($\mathcal{L}\setminus\mathcal{T}$ marked in gray), averaged
over windows. \textbf{(c)} Illustration of the ordinary least squares (OLS) fit between measured OTPM and probe gain. Based on the OTPM of an unseen language pair, we can predict the probe gain as shown with the orange marker (input OTPM of $0.3$ yields a predicted gain $\hat{\Delta}$ of ${\sim}6.9$\,pt).
  \label{fig:diagnostic_pipeline}
}

\end{figure*}

\section{Methodology}

\subsection{Spoken language diarization}
Given a speech recording and a set $\mathcal{T}$ of target languages for which we want to build the SLD system, the objective is to output time-stamped segments, each labelled with one of the languages in $\mathcal{T}$. 
A cascaded SLD system is composed of three stages: (1) voice activity detection (VAD) to identify speech regions, (2) segmentation of speech utterances into short overlapping windows followed by feature extraction from a pre-trained speech model (often a language identification posterior $p_w$ as shown in Fig.~\ref{fig:diagnostic_pipeline}(a)), and (3) clusters the windows to form the corresponding language clusters.
Given labels for each cluster, evaluation is performed using Language Diarization Error Rate (LDER) metric. 

\begin{equation}
  \mathrm{LDER} \;=\; \frac{T_{\text{miss}} + T_{\text{FA}} + T_{\text{conf}}}{T_{\text{ref}}},
  \label{eq:lder}
\end{equation}
LDER is analogous to Diarization Error Rate in speaker diarization context which is computed as the summation of missed ($T_\text{miss}$), falsely detected ($T_\text{FA}$) and wrongly labelled ($T_\text{conf}$) speech time divided by the total reference speech time.\citep{ryant2019dihard} 
In addition to LDER, we observe that for some multilingual speech recordings the output of the clustering model provides a single language label indicating that the model is not able to diarize the recording.
We refer to this phenomenon as \emph{collapse} and report these values  alongside LDER.

\subsection{Off-target LID posterior mass}
\label{sec:offm}

Let the language identification posterior $p_{w}\in \mathbf{R}^{|\mathcal{L}|}$ of the pre-trained model on a window $w$ support a fixed set of $\mathcal{L}$ languages (for instance, Whisper supports 99 languages).
\emph{Target} languages $\mathcal{T} \subseteq \mathcal{L}$ are the languages the SLD model must diarize, fixed by the downstream task; \emph{non-target} languages form the complement $\mathcal{L}\setminus\mathcal{T}$, i.e., every other language the model supports (red and gray bars, respectively, in Fig.~\ref{fig:diagnostic_pipeline}(b)).

\noindent\textbf{Definition.} Let the corpus consist of speech utterances $f \in F$, and let $W_f$ be the windows of utterance $f$ that is used as input to the clustering stage (Sec.~\ref{sec:setup}). 
Off-Target Posterior Mass (OTPM) is defined as the posterior mass the pre-trained model assigns to non-target languages, averaged over the windows of each utterance:
\begin{equation}
    \text{OTPM} \;=\; \frac{1}{|F|}\sum_{f\in F}\frac{1}{|W_f|}\sum_{w\in W_f}
    \;\sum_{l \in \mathcal{L}\setminus\mathcal{T}} p_{w}(l),
    \label{eq:off_target_corpus}
\end{equation}
where $p_{w}(l)$ is the posterior probability the model assigns to language $l$ on window $w$.
For example, when performing SLD with Hindi and English we have $\mathcal{T}=\{\texttt{hi},\texttt{en}\}$, if a window has $p(\texttt{hi}){=}0.55$ and $p(\texttt{en}){=}0.05$, then the OTPM for this window would be $0.40$ i.e. the posterior mass assigned to non-target languages.
OTPM lies in the range $[0,1]$, with values close to zero indicating that the pre-trained posterior probabilities are assigned entirely to target languages; whereas values close to 1 imply almost all of the posterior probability is assigned to non-target languages ($\mathcal{L} \setminus \mathcal{T}$).
Based on the definition in Eq.~\eqref{eq:off_target_corpus}, OTPM depends on the downstream corpus, the target set of languages and the pre-trained speech model, but does not need explicit language labels (only requires the set of languages in the corpus) and does not require clustering/fine-tuning.
To get an accurate value of OTPM, it must be computed on the windows of the recordings that are representative of the corpus.

\subsection{Probe gain}
SLD performance in a cascaded system can be improved through fine-tuning the posterior probability distributions produced by the pre-trained speech models for individual speech windows. 
We refer to a classifier trained on target-language data on top of the frozen pre-trained model as a \emph{probe}, wherein the pre-trained model is frozen.
\textbf{Probe Gain} refers to the improvement in LDER obtained when clustering the probe's outputs instead of the pre-trained posterior probabilities.
\begin{equation}
  \Delta \;=\; \mathrm{LDER}_\texttt{Pretrain} \;-\; \mathrm{LDER}_\texttt{Probe},
  \label{eq:probe_gain}
\end{equation}
where $\mathrm{LDER}_\texttt{Probe}$ refers to LDER from clustering with posterior probabilities with probe model and $\mathrm{LDER}_\texttt{Pretrain}$ refers to LDER on test set from using posterior probabilities from pre-trained model.
A larger $\Delta$ indicates that the pre-trained model posteriors do not adequately capture the distinction between the target languages.

\noindent \textbf{Probe Variants.} In this work, we test OTPM with two probe variants, both logistic regressions on frozen features: (1) the \emph{posterior probe} is trained on top of pre-trained LID posterior $p_w$ (99 dimensions for Whisper \texttt{base}, 100 for large-v3); (2) the \emph{encoder probe} uses standardized final-layer encoder output as input, mean-pooled over the temporal window (resulting in 512 and 1280 dimension vector for Whisper base and large-v3 respectively).
In both cases, the probe's class probabilities replace the pre-trained posterior as the clustering input.

\subsection{Hypothesis}
\label{sec:hypothesis}
We hypothesize that Off-target posterior mass (OTPM) calculated for a corpus of utterances can be used as a proxy for the fine-tuning probe gain ($\Delta$) that can be obtained for SLD of the target languages in the corpus.
This hypothesis is intuitive as larger OTPM indicates that fine-tuning would re-distribute the posterior mass from non-target languages to target-languages, thereby leading to better diarization.
Hence, we would expect that larger the OTPM value, higher the expected gain from fine-tuned SLD probe.
To establish this relationship, we build an ordinary least squares model (as illustrated in Fig.~\ref{fig:diagnostic_pipeline}(c)) to predict SLD probe gain from OTPM for a set of training SLD language pairs and tested on unseen corpora and languages.

\section{Experimental setup}
\label{sec:setup}
\noindent\textbf{Data.} 
Firstly, hypothesis stated in Sec.~\ref{sec:hypothesis} was tested on synthetically generated bilingual corpora from language pairs sampled from FLEURS dataset.\citep{conneau2023fleurs}
FLEURS\citep{conneau2023fleurs} is a multilingual read-speech corpus of monolingual utterances, which lets us choose the target languages for diarization enabling testing across widely and less commonly spoken languages. 
We fit on one training set of 32 language pairs and test the fits on a held-out set of 19 unseen test pairs.
The 32 training pairs are Yoruba--Amharic, Swahili--Hausa, Bengali--Tamil, Vietnamese--Thai, French--German, Japanese--Korean, Spanish--Portuguese, Filipino--Indonesian, Assamese--Nepali, Persian--Pashto, Marathi--Gujarati, Malay--Filipino, Vietnamese--Khmer, Dutch--Swedish, Arabic--Hebrew, Italian--Romanian, Russian--Ukrainian, Polish--Czech, Bulgarian--Macedonian, Lithuanian--Latvian, Finnish--Estonian, Turkish--Azerbaijani, Kazakh--Uzbek, Kannada--Telugu, Punjabi--Sindhi, Danish--Norwegian, Afrikaans--Luxembourgish, Catalan--Galician, Shona--Lingala, Mandarin--Burmese, Georgian--Armenian, and Javanese--M\=aori.
Synthetic data for each language pair was generated from concatenation of 10 test-split utterances per language with 0.3\,s gaps; each language pair comprises 45 utterances unless one of the languages data size was shorter than the required duration. 
Following this, OTPM was computed for each language pair resulting in a range of $0.015$ to $0.978$ across the 32 training pairs.
\par 
\noindent To test the generalizability of hypothesis, validation on real-world multilingual speech corpora was performed: (1) SEAME\citep{lyu2010seame} Mandarin--English (30 development recordings, 21.0\,h of audio, 11.1\,h of reference speech) (2) VoxPopuli\citep{wang2021voxpopuli} German and English (40 full European Parliament sessions whose aligned speech is at least 90\% in those two languages; 36.2\,h of audio, 9.0\,h of aligned reference speech) and (3) DISPLACE 2024,\citep{displace2024} a far-field multi-party conversational benchmark with three target languages (English, Hindi, Malayalam; 32 evaluation files, $17.2$\,h of reference speech under three language labels). 
Across all corpora, evaluation was performed on unseen recordings, with training performed on 150 SEAME, 60 VoxPopuli and 34 DISPLACE recordings.
\par
\noindent\textbf{Diarization pipeline.} 
All experiments performed throughout the manuscript share a common cascaded pipeline. 
For the real corpora, Silero VAD\citep{silero} filters non-speech segments.
For synthetically generated data, each recording is treated as a single speech region. 
Each speech region is then split into 10\,s windows at a 0.4\,s hop. If a region is shorter than 20\,s, it is merged into a single window.
Window level features are then extracted from pre-trained/fine-tuned models followed by cosine agglomerative hierarchical clustering (average linkage, threshold from a two-component Gaussian-mixture calibration of the similarity score and offset $0.015$).
Additionally, one VBx resegmentation pass\citep{landini2022bayesian} ($\mathrm{Fa}{=}1.5$, $\mathrm{Fb}{=}2$) was performed. 
LDER is computed using \texttt{dscore}\citep{ryant2019dihard} with no collar and includes overlapping speech in scoring.

\noindent\textbf{Models.} Whisper \texttt{base} (74M) is the pre-trained model used for all main results in this work.
Synthetic data experiments were repeated with Whisper \texttt{large-v3} (1.55B) in Sec.~\ref{sec:largev3} to demonstrate generalizability of observed results.

\noindent\textbf{Statistical analysis.} We fit ordinary least squares model between SLD probe gain and OTPM ($\hat{\Delta} = \alpha + \beta\,\text{OTPM}$) and report the slope with a 95\% residual-bootstrap CI (10k resamples; a pairs bootstrap for the $N{=}50$ reference runs), in-sample $R^2$ and leave-one-out $Q^2$.
Correlation between OTPM and probe gain was tested with one-sided Spearman $\rho$, with a stratified $\rho$ (the file-weighted mean of the per-corpus values) across the files of real-world corpora and permuted for within corpora. 
To test statistically significant differences between LDER of pre-trained and fine-tuned models, $p$-values from centered paired bootstrap across evaluation recordings with 10k resamples were compared.
\begin{figure*}[t]
  \centering
  \includegraphics{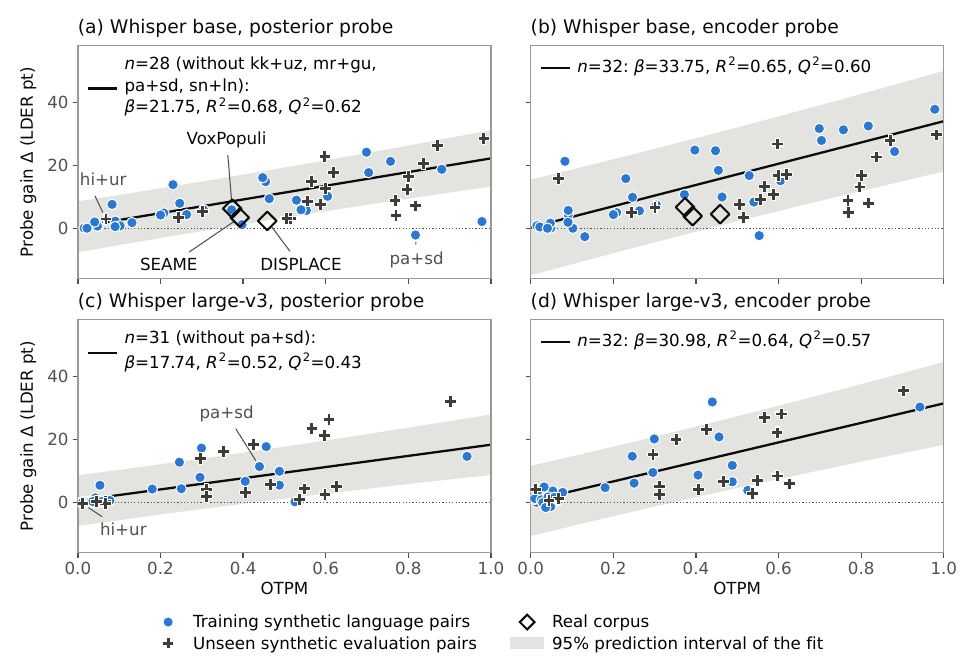}
    \caption{Probe gain vs.\ OTPM at $N{=}400$ training utterances per language, (a) Whisper \texttt{base} posterior-probe, (b) Whisper \texttt{base} encoder-probe, (c) Whisper large-v3 posterior  probe and (d) Whisper large encoder probe. Blue Circles indicate the 32 training synthetic language pairs. Plus markers indicate the 19 unseen synthetic evaluation pairs.
    Black line shows the OLS fit on the training pairs and each panel reports its slope $\beta$, in-sample $R^2$ and leave-one-out $Q^2$ (95\% CIs of $\beta$ ). Shaded regions in gray mark 95\% prediction interval. The encoder-probe fits use all 32 training pairs. The posterior-probe fits leave out the training pairs which produce collapsed posteriors.
    Labelled examples: pa+sd (Punjabi--Sindhi) and hi+ur (Hindi--Urdu); kk+uz, mr+gu and sn+ln are Kazakh--Uzbek, Marathi--Gujarati and Shona--Lingala. Markers in diamonds in subplot (a, b) indicate real corpora, whose probes are trained on in-domain recordings.}
  \label{fig:misalignment}
\end{figure*}
\section{Results}
\label{sec:results}

To test the hypothesis stated in Sec.~\ref{sec:hypothesis}, firstly we train an OLS model with 32 training synthetic bilingual corpora using the OTPM measured on the corpora and the probe gain ($\Delta$) obtained from fine-tuning with the corpora.
This OLS model is then used to predict the probe gain on an unseen set of 19 synthetically generated bilingual language pairs.
Then, we test whether the OLS models predictions translate to real-world corpora through comparing the predicted probe gain with OTPM to relevant baselines.
Finally, we show that the observed fit between OTPM and probe gain is generalizable across models through replicating synthetic data experiments with Whisper large-v3 model.
\subsection{OTPM is indicative of SLD probe gain}
\label{sec:regression}

OLS fit to predict probe gain (for encoder and posterior probe) from observed OTPM values for synthetic language pairs from FLEURS dataset has a positive slope as shown in Fig.~\ref{fig:misalignment}.
Fine-tuned probes (encoder probe, posterior probe) are trained on each language pair with $N{=}400$ utterances, with LDER and OTPM reported on test utterances. 
This observation aligns with the stated hypothesis that fine-tuning probes are able to provide improvements in diarization ability through directing the posterior mass on non-target languages onto target languages.
For the encoder probe, we observe a large positive slope $\beta{=}33.75$ (95\% CI $[25.1, 42.4]$) alongside $R^2{=}0.65$, $Q^2{=}0.60$ whereas posterior probe results indicate a more moderate positive slope with $\beta{=}13.61$ $[6.6, 20.6]$ and $R^2{=}0.31$, $Q^2{=}0.14$. 
Poor fit for posterior probe is attributed to three language pairs, for which the fit predicts far more gain than observed: Punjabi--Sindhi ($\text{OTPM}{=}0.82$, $\Delta_{\text{True}}{=}{-}2.20$, $\Delta_{\text{Pred}}{=}13.1$\,pt), Shona--Lingala ($\text{OTPM}{=}0.98$, $\Delta_{\text{True}}{=}2.09$, $\Delta_{\text{Pred}}{=}15.3$) and Marathi--Gujarati ($\text{OTPM}{=}0.40$, $\Delta_{\text{True}}{=}1.12$, $\Delta_{\text{Pred}}{=}7.4$), while the encoder probe obtains LDER gains of $32.45$, $37.74$ and $24.78$ on the same pairs.  
For these three pairs, clustering with posterior probe leads to \emph{collapse}, a single cluster rather than two distinct clusters for the two languages resulting in minimal probe gains. 
Additional discussion related to clustering collapse into single cluster is provided in Sec.~\ref{sec:collapse}.
Re-computing without the four collapsed pairs (Sec.~\ref{sec:collapse}) provides a steeper slope for posterior probe, $\beta{=}21.75$ $[16.4, 27.4]$, and a much better fit ($R^2{=}0.68$, $Q^2{=}0.62$, $n{=}28$), as shown in Fig.~\ref{fig:misalignment}.

\subsection{Generalization of OLS fit to unseen pairs}
\label{sec:fresh}

While the OLS fit on the 32 training pairs is able to show linear relationship between OTPM and SLD probe gain, we test whether we are able to predict probe gains on unseen language pairs.
This is tested by generating 19 additional unseen language pair corpora from FLEURS dataset.
The observed probe gains for each of the 19 unseen pairs are plotted in Fig.~\ref{fig:misalignment} both for encoder and posterior probe settings.  
Firstly, the Spearman correlation between OTPM and probe gain on the unseen pairs is positive for both probes: $\rho{=}0.68$ (one-sided $p{=}0.001$) for the posterior probe on the 18 pairs not flagged for collapse, and $\rho{=}0.56$ (one-sided $p{=}0.006$) for the encoder probe. 
While the correlation between OTPM and probe gains is high, the absolute magnitude of predictions indicate mixed results.
While posterior-probe fit consistently outperformed a mean prediction baseline (mean squared error, MSE, $44.74$ vs.\ $92.19$; out-of-sample $R^2{=}0.52$), the encoder-probe fit performs worse than constant mean prediction ($120.87$ vs.\ $64.32$; $R^2{=}-0.88$).

\subsection{Impact of training-set size on SLD probe gain}
\label{sec:trainsize}
To test whether the positive slope between OTPM and probe gain arises primarily from size of training data, we perform OLS fits for training set size $N\in\{50, 100,200,300,400\}$ while holding evaluation utterances and clustering algorithm fixed.
Results are reported in Table~\ref{tab:trainsize}, and for all training set sizes a positive slope is observed on OLS fit for both encoder probe and posterior probe. 
While the encoder probe slope across different training set sizes remains constrained to a small range $33.4$--$33.7$, posterior probe OLS slope increases from $5.4$--$13.6$.
Within posterior probe slopes, there is stark increase in slope from $N{=}50$ to $N{=}200$ followed by a plateau.
We observe that this increase in slope is primarily arising from clustering collapse of Yoruba--Amharic languages into a single cluster, leading to noticeably lower SLD probe gain in smaller training sets.
\begin{table}[t]
  \centering
  \caption{Slope $\beta$ of probe gain on OTPM [95\% CI] by training-set size $N$ (utterances per language). $Q^2$ at $N{=}400$ is reported for 32 FLEURS training pairs. Non-coll. indicates OLS fit with language pairs whose posterior probe do not collapse at that training size $N$ (non-collapsed language pairs respectively are $n{=}26, 28, 29, 29, 29$). Tests of the $N{\times}\text{OTPM}$ interaction: $p{=}0.035$, $0.020$, $0.69$ (rows 1--3 respectively). $^\dagger$Probes trained on reserved test-split clips.}
  \label{tab:trainsize}
  \footnotesize
  \setlength{\tabcolsep}{3pt}
  \begin{tabular}{lcccccc}
    \toprule
    \textbf{Probe} & $N{=}50^\dagger$ & $100$ & $200$ & $300$ & $400$ & $Q^2$ \\
    \midrule
    Posterior, all     & $5.4$ {\scriptsize$[-2.9, 15.9]$} & $8.2$ {\scriptsize$[1.9, 14.9]$} & $12.2$ {\scriptsize$[5.7, 18.8]$} & $13.2$ {\scriptsize$[6.5, 20.1]$} & $13.6$ {\scriptsize$[6.6, 20.6]$} & $0.14$ \\
    Posterior, non-coll. & $18.8$ {\scriptsize$[11.1, 25.8]$} & $16.7$ {\scriptsize$[10.1, 23.2]$} & $20.6$ {\scriptsize$[15.3, 26.1]$} & $22.1$ {\scriptsize$[17.1, 27.5]$} & $22.8$ {\scriptsize$[17.6, 28.1]$} & $0.67$ \\
    Encoder, all       & $36.8$ {\scriptsize$[29.1, 43.5]$} & $33.4$ {\scriptsize$[24.8, 41.9]$} & $33.6$ {\scriptsize$[24.7, 42.5]$} & $33.7$ {\scriptsize$[24.9, 42.3]$} & $33.7$ {\scriptsize$[25.1, 42.4]$} & $0.60$ \\
    \bottomrule
  \end{tabular}
\end{table}

\subsection{Translation of OLS fit to real-world speech}
\label{sec:realworld}

While experiments using FLEURS dataset on synthetically generated data establish the relationship between OTPM and SLD probe gain, we test the extent to which the OLS model is able to translate to real-world speech corpora.
Results for OTPM values along with the true SLD probe gains and predicted gains on three real-world corpora, both for Posterior probe and Encoder probe are reported in Table~\ref{tab:realcorpora}.
Firstly, we observe that OTPM predictions of probe gains are noticeably higher compared to actual gains.
This could be owing to the mismatch between the manner in which synthetic data was constructed compared to real world multilingual speech corpora. 
While the synthetic data is composed of utterances that contains equal proportions of both languages in a given recording, often real world corpora contain recordings that are primarily single language with code-switched or code-mixed second language.
This results in a sim-to-real distribution shift which leads to our OLS model over-predicting the expected gains.

Although OTPM over-predicts the gains on real corpora, OTPM is a better predictor of probe gains compared to using LDER of pre-trained model in the OLS model (posterior-probe MSE $42.6$ vs. LDER MSE $428$).
Within each corpus, OTPM orders recordings by probe gain as indicated by the per-recording Spearman $\rho$, averaged over corpora, is $0.23$ for the encoder probe and $0.27$ for the posterior probe (Table~\ref{tab:realcorpora}), with highest correlation on VoxPopuli dataset.

\begin{table*}[t]
  \centering
  \caption{Results on the three real corpora for Whisper \texttt{base}.
  $\Delta_{\text{post}}$ and $\Delta_{enc}$ refer to the true LDER probe gains measured on posterior and encoder probes for the respective corpora with p-values indicating whether the gain was significantly above zero. $\Delta_{\text{post}}^{\text{OTPM}}$ and $\Delta_{\text{post}}^{\text{LDER}}$ refer to the posterior probe gains predicted from OLS models trained using OTPM and baseline LDER respectively.
    Within-corpus $\rho$ indicates how well OTPM orders the recordings of each corpus by corresponding probe gains.
    SEAME, VoxPopuli and DISPLACE datasets have 30, 40 and 32 test recordings respectively with collapsed recordings  for (pre-trained / posterior / encoder) for each dataset being SEAME 13/14/16, VoxPopuli 8/11/9 and DISPLACE 11/14/10. Collapsed recordings are ignored during computation of Spearman correlation. Average over corpora reports the average Spearman correlation for recordings across all three corpora combined.}
  \label{tab:realcorpora}
  \footnotesize
  \setlength{\tabcolsep}{2.2pt}         
    \begin{tabular}{lcc|cc|cc|cc}
    \toprule
    & & & & & \multicolumn{2}{c}{\textbf{Predicted} $\Delta_{\text{post}}$}
      & \multicolumn{2}{c}{\textbf{Within-corpus} $\rho$} \\
    \cmidrule(lr){6-7}\cmidrule(lr){8-9}
    \textbf{Corpus} & \textbf{OTPM} & \textbf{LDER}$_{\texttt{pre}}$ &
      \textbf{$\Delta_{\text{post}}$} ($p$) & \textbf{$\Delta_{\text{enc}}$} ($p$) &
      $\Delta_{\text{post}}^{\text{OTPM}}$ & $\Delta_{\text{post}}^{\text{LDER}}$ & Encoder & Posterior \\
    \midrule
    SEAME      & 0.392 & 68.61 & $+3.34$ (0.003) & $+3.81$ (0.002)  & $8.93$  & $33.77$ & $0.06$  & $0.33$ \\
    VoxPopuli  & 0.374 & 25.82 & $+6.20$ (0.004) & $+6.62$ (0.003)  & $8.53$  & $9.65$  & $0.43$  & $0.34$ \\
    DISPLACE   & 0.458 & 37.05 & $+2.28$ (0.17)  & $+4.44$ (0.0025) & $10.37$ & $15.98$ & $0.27$  & $0.15$ \\
    \midrule
    \multicolumn{7}{l}{Average over corpora (one-sided $p$)} & $0.23$ (0.0015) & $0.27$ (0.012) \\
    \bottomrule
  \end{tabular}
\end{table*}

\subsection{Comparison of OTPM to baseline metric}
OTPM is one metric that could be used to predict SLD probe gains.
To demonstrate its effectiveness, both in terms of ease of computation and predictability of probe gains, we compare its performance to other plausible metrics.
\textbf{Baseline LDER} as a metric refers to using LDER obtained from clustering with pre-trained model posteriors to predict the SLD probe gain, whose OLS equation would be: $\hat{\Delta} = \alpha + \beta\,\text{LDER}_{\text{Pretrain}}$. 
Within FLEURS baseline LDER provides a better fit than OTPM ($Q^2{=}0.72$ for posterior probe and $Q^2{=}0.81$ for encoder probe).
This could be owing to baseline LDER computation accounting for confusion between the two target languages explicitly, while OTPM does not. 
However OTPM has two distinct advantages, firstly OTPM is computed in a training-free and label-free manner without requiring language level annotations, whereas pre-trained LDER requires recording level labels alongside clustering.
Secondly, when applied to real-world corpus we observe that the baseline LDER OLS fit over-predicts the probe gain on every dataset compared to OTPM as reported in Table~\ref{tab:realcorpora}.

\subsection{Generalization of probe gain fit with Whisper large-v3}
\label{sec:largev3}
\begin{table}[t]
  \centering
  \caption{Whisper \texttt{base} vs.\ large-v3, $N{=}400$. Fit on the 32 training pairs for posterior probe ($n$ indicates the language pairs that did not have a posterior collapse during clustering, $\beta$ refers to the slope of fit along with 95\% CI and $Q^2$). Test on the 19 unseen language pairs: Spearman $\rho_{\text{test}}$ and out-of-sample $R^2_{\text{test}}$ against a constant baseline ($^{*}$ indicates 95\% bootstrap interval of the improvement excludes zero).}
  \label{tab:largev3}
  \footnotesize
  \setlength{\tabcolsep}{3.5pt}
  \begin{tabular}{l rlrrr c rlrrr}
    \toprule
    & \multicolumn{5}{c}{\textbf{Posterior probe}} & & \multicolumn{5}{c}{\textbf{Encoder probe}} \\
    \cmidrule(lr){2-6}\cmidrule(lr){8-12}
    & $n$ & $\beta$ {\scriptsize[95\% CI]} & $Q^2$ & $\rho_{\text{test}}$ & $R^2_{\text{test}}$
    & & $n$ & $\beta$ {\scriptsize[95\% CI]} & $Q^2$ & $\rho_{\text{test}}$ & $R^2_{\text{test}}$ \\
    \midrule
    \texttt{base} & 28 & $21.8$ {\scriptsize$[16.4, 27.4]$} & $0.62$ & $0.68$ & $0.52$
                  & & 32 & $33.7$ {\scriptsize$[25.1, 42.4]$} & $0.60$ & $0.56$ & $-0.88$ \\
    large-v3      & 31 & $17.7$ {\scriptsize$[12.2, 24.5]$} & $0.43$ & $0.65$ & $0.50^{*}$
                  & & 32 & $31.0$ {\scriptsize$[23.2, 39.9]$} & $0.57$ & $0.69$ & $0.53$ \\
    \bottomrule
  \end{tabular}
\end{table}
In addition to Whisper-base model, we repeat the synthetic-data experiments with Whisper large-v3 (1.55B parameters) to show generalization across models.
From the results in Table~\ref{tab:largev3}, the OLS fit of OTPM on probe gain with 32 training pairs gives a leave-one-out $Q^2{=}0.43$ for the posterior probe ($n{=}31$ non-collapsed pairs) and $Q^2{=}0.57$ for the encoder probe.
On the test pairs, the Spearman correlation between OTPM and probe gain is significantly positive: one-sided $p{=}0.002$ for the posterior probe on the 18 test pairs that do not collapse, and $p{=}0.0005$ for the encoder probe.
Unlike with \texttt{base}, the frozen posterior-probe fit also beats the constant prediction baseline with a confident margin (95\% interval of the reduction in mean squared error $[5.9, 151.9]$).
These observations validate our hypothesis that OTPM can act as a training-free metric to forecast SLD probe gains on unseen corpora.
As Whisper large-v3 is much larger compared to Whisper-base and consequently is able to better model multilingual speech utterances, the OTPM range observed across different language pairs is smaller than Whisper-base.
The mean OTPM values across the test and evaluation pairs for Whisper-base is $0.46$ while that of Whisper large-v3 is much smaller at $0.26$.
Consequently, Whisper large-v3 SLD probe provides lower mean probe gain of $5.8$ LDER compared to $8.7$ for Whisper-base.

\section{Discussion}
\label{sec:discussion}

\textbf{Impact of pre-training hours on OTPM.}
While OTPM is able to track probe gains, one of the possible explanations for this could be that OTPM is indicator of the number of pre-training hours corresponding to a specific language, larger the pre-training data better the models ability at recognizing the language.
Spearman correlation of $-0.77$ between $\log_{10}$ of Whisper's speech-recognition training hours ($H_{\text{Pretrain}}$)\citep{radford2023whisper} and observed OTPM for the language indicate that the number of pre-training hours have a noticeable impact on OTPM observed.
When using $H_{\text{Pretrain}}$ to forecast SLD probe gains instead of OTPM, we observe that for posterior probe provides a Spearman $\rho$ of $0.9$ for training pairs ($\rho_{OTPM}^{\text{Posterior}}{=}0.84$), and a Spearman $\rho$ of $0.83$ for training pairs ($\rho_{OTPM}^{\text{Encoder}}{=}0.74$).
However, on the unseen 19 language pairs pre-training hours noticeably underperforms OTPM with $\rho_{OTPM}^{\text{Posterior}}{=}0.68$ vs $\rho_{\text{Pretrain Hours}}^{\text{Posterior}}{=}0.32$, and $\rho_{OTPM}^{\text{Encoder}}{=}0.56$ vs $\rho_{\text{Pretrain Hours}}^{\text{Encoder}}{=}0.31$.
This result indicates that OTPM is able to capture information specific to the dataset and recording conditions beyond number of pre-training hours of the languages, indicated by the better correlation values.

\noindent \textbf{Utility of OTPM as a metric for SLD modeling.} 
The primary utility of OTPM arises from the ability to guide whether a given dataset is useful as a fine-tuning dataset prior to annotating the dataset for language labels. 
Based on the value of the OTPM, a user can make an informed decision of whether the obtained probe gain justifies annotation cost and corresponding fine-tuning compute requirements. 
Within a real corpus, OTPM points to the recordings where a probe helps most, though at a stratified $\rho$ of about $0.25$ it is a weak guide for any single recording (Sec.~\ref{sec:realworld}). 

\noindent \textbf{Posterior Probe Collapse.}\label{sec:collapse}
In some instances during clustering of the posterior probabilities to obtain language diarization intervals, we observe that the results return a single language despite the recording containing two distinct languages.
While no encoder probe model encounters this for the 32 training language pairs, six pairs for posterior probe at $N{=}50$, four at $N{=}100$, and on three pairs at $N{=}200$ on
(Marathi--Gujarati, Punjabi--Sindhi, Shona--Lingala) result in posterior collapse. 
While the collapse for some language pairs (Kazakh--Uzbek, Kannada--Telugu, Yoruba--Amharic) was resolved at large training sample sizes, the rest persists across training sample sizes, consistent with both target language posteriors leaking onto the same off-target languages in similar proportions, which prevents the posterior probe from being able to separate the languages.

\section{Limitations}
\label{sec:limits}

Proposed OTPM model is able to provide robust results on synthetically generated data and generalize to unseen languages in this setting but the OLS model trained on synthetic data shows limited generalizability to real-world data.
We hypothesize that this gap primarily arises from our synthetically generated data containing equal proportions of both languages in the recordings, whereas real-world recordings contain majority of a single language with parts of second language in between. 
Hence, a more realistic set of synthetically generated data could potentially generalize better to real-world corpora.
Additionally, while OTPM accounts for the posterior probabilities assigned to non-target languages in SLD, it does not explicitly model language confusion between the target languages (which are non-negligible for languages in the same family\citep{kommineni2025can}).
This phenomenon is illustrated with two examples, firstly in Turkish--Azerbaijani has $\text{OTPM}{=}0.08$ but has probe gains of $+7.54$ (posterior) and $+21.23$\,pt (encoder) against predictions of $+2.2$ and $+3.0$, and Hindi--Urdu shows similar observations with OTPM $0.07$ but encoder-probe gain $+15.71$ against $+2.49$ predicted.
This could be another factor that explains the limited generalizability of the proposed OTPM metric. 

\section{Conclusion}
\label{sec:conclusion}

We proposed off-target LID posterior mass (OTPM), the proportion of a pre-trained model's LID posterior that falls outside the target languages, as a way to forecast SLD probe gains  from unlabelled audio.
Results on synthetic data indicate that OTPM is able to generalize on unseen language pairs for predictions of probe gain, both on Whisper base and Whisper large-v3 models.
Additionally, through ablations on the training-set size and comparison to baselines we show that OTPM is able to capture meaningful information pertaining to probe gains.
While results on real-world data provide positive correlation for recording level probe gain, on corpus level data we see that OTPM suffers from lack of robustness owing to gap between synthetic and real-world data characteristics and lack of accounting for language confusion within target languages.
Future works should attempt to account for these limitations so as to develop robust diagnostic metrics for SLD modeling.

\begin{acknowledgments}
\noindent
Large language model assistance was used for writing code, generating plots and formatting tables.
The core research ideas and scientific content are solely the authors', and LLMs were not used.
All text and code included as a part of the manuscript was verified by the authors.
\end{acknowledgments}

\section*{Author Declarations}
\noindent
The authors have no conflicts to disclose.

\section*{Data Availability}
\noindent
All datasets used in this work are available publicly. 
FLEURS and VoxPopuli are available from their original releases.\citep{conneau2023fleurs, wang2021voxpopuli} 
SEAME and DISPLACE 2024 corpora are available from their original distributors under their respective licenses.\citep{lyu2010seame, displace2024}
The code and per-pair results that support the findings of this study will be open-sourced upon manuscript acceptance.

\bibliography{refs}

\end{document}